\documentclass[a4paper,11pt]{article}
\usepackage{pos}
\usepackage{mathtools}
\usepackage{todonotes}
\usepackage{tikz}
\usepackage{caption}
\usepackage{subcaption}
\usetikzlibrary{decorations,decorations.markings}
\usetikzlibrary{decorations.shapes}

\newcommand{\Cstar}{C${}^*$~}

\title{Baryon masses from full QCD+QED${}_\text{C}$ simulations}

\author[a]{Lucius Bushnaq}
\author[b]{Isabel Campos}
\author[c]{Marco Catillo}
\author[d]{Alessandro Cotellucci}
\author*[e,f,g]{Madeleine Dale}
\author[a]{Patrick Fritzsch}
\author[d,h]{Jens Lücke}
\author[c]{Marina Krsti\'{c} Marinkovi\'{c}}
\author[d,h]{Agostino Patella}
\author[e,f]{Nazario Tantalo}

\affiliation[a]{School of Mathematics, Trinity College Dublin, Dublin 2, Ireland
}
\affiliation[b]{
Instituto de Física de Cantabria \& IFCA-CSIC, Avda. de Los Castros s/n, 39005 Santander, Spain
}
\affiliation[c]{Institut für Theoretische Physik, ETH Zürich, Wolfgang-Pauli-Str. 27, 8093 Zürich, Switzerland
}
\affiliation[d]{Humboldt Universität zu Berlin, Institut für Physik {\&} IRIS Adlershof, Zum Gro{\ss}en Windkanal 6, 12489
Berlin, Germany
}
\affiliation[e]{Università degli Studi di Roma "Tor Vergata",
 Via Cracovia 50, 00133 Rome, Italy}

\affiliation[f]{INFN, Sezione di Tor Vergata,
Via della Ricerca Scientifica 1, 00133 Rome, Italy}

\affiliation[g]{University of Cyprus, Department of Physics, 1 Panepistimiou Street, 2109 Aglantzia, Nicosia, Cyprus}

\affiliation[h]{DESY, Platanenallee 6, D-15738 Zeuthen, Germany
}

\emailAdd{bushnaql@tcd.ie}
\emailAdd{isabel.campos@csic.es}
\emailAdd{mcatillo@ethz.ch}
\emailAdd{alessandro.cotellucci@physik.hu-berlin.de}
\emailAdd{m.dale@stimulate-ejd.eu}
\emailAdd{fritzscp@tcd.ie}
\emailAdd{jens.luecke@hu-berlin.de}
\emailAdd{marinama@phys.ethz.ch}
\emailAdd{agostino.patella@physik.hu-berlin.de}
\emailAdd{nazario.tantalo@roma2.infn.it}

\abstract{In these proceedings we present preliminary results for the masses of the proton, neutron and $\Omega^-$ baryons obtained from QCD+QED lattice simulations performed with four dynamical quarks using C$^*$ boundary conditions. 
These results are part of the ongoing effort of the RC${}^*$ collaboration discussed in the companion proceedings~\cite{bushnaq2021update}, and have been obtained on a single ensemble in which the renormalised electromagnetic coupling is $\alpha_{\text{em}}\sim 0.04$, the physical volume is $L\sim 1.7$~fm and the masses of the four dynamical quarks have been tuned at the $U$--spin symmetric point $m_d=m_s$. We demonstrate on this unphysical ensemble that baryon masses can be calculated with satisfactory precision when including QED without the need for gauge--fixing and perturbation theory. This makes us confident in the effectiveness of the strategy presented here also in the case of simulations closer to the physical point.
\vspace{5mm}
\begin{center}
\includegraphics[width=0.2\linewidth]{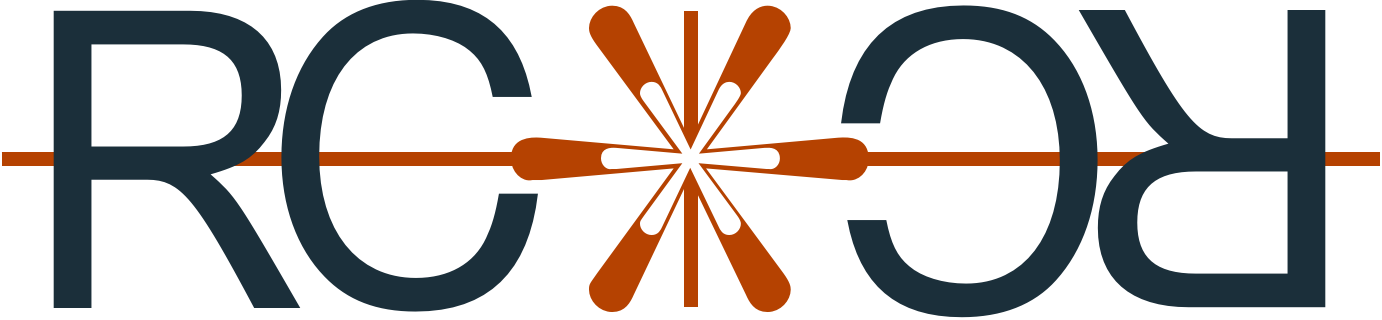}
\end{center}}

\FullConference{%
 The 38th International Symposium on Lattice Field Theory, LATTICE2021
  26th-30th July, 2021
  Zoom/Gather@Massachusetts Institute of Technology
}

\begin{document}
\maketitle

\section{Introduction}
The statistical error in lattice QCD for certain quantities, such as for low-lying baryon masses, is currently reduced to levels such that it is of a similar magnitude to the systematic error due to neglecting QED effects.
Pushing forward for increased precision therefore requires the inclusion of QED if accuracy is to be maintained. 

One complication in the formulation of QED on the lattice is that Gauss' Law dictates that, on a torus with periodic boundary conditions, only states with net-zero electric charge belong to the physical Hilbert space. 
A naive local gauge--fixing results in unconstrained global zero-modes. The principle behind QED$_L$, the more common formulation of lattice QED, 
is to decouple zero-modes from gauge field dynamics through enforcing the constraint $\int_{L^3} d^3x A_\mu(t,\mathbf{x}) = 0$. A disadvantage of this approach is that this constraint is non-local.
 Therefore, it is not guaranteed that properties such as renormalisability will hold; they must be proven for individual observables, although it is relatively simple to address these issues at $\mathcal{O}(\alpha)$, and in fact it has been proven that at this order these properties hold for the spectrum. 
 
The  QED$_L$ formulation has been used in the reference work on the baryon spectrum by the BMW collaboration~\cite{bmw_2015}. 
In this work, we will use the \Cstar formulation~\cite{kronfeld1991,tantalo_2016} of QED on the lattice, which is based upon enforcing $C$-parity boundary conditions along the spatial directions for all the fields. 
This results in a spatially anti-periodic $U(1)$ gauge field, meaning that spatial zero-modes sum to zero. 
As this approach is totally local, renormalisability is guaranteed. Thus the spectra of electrically--charged states may be calculated without perturbation theory or gauge--fixing. 

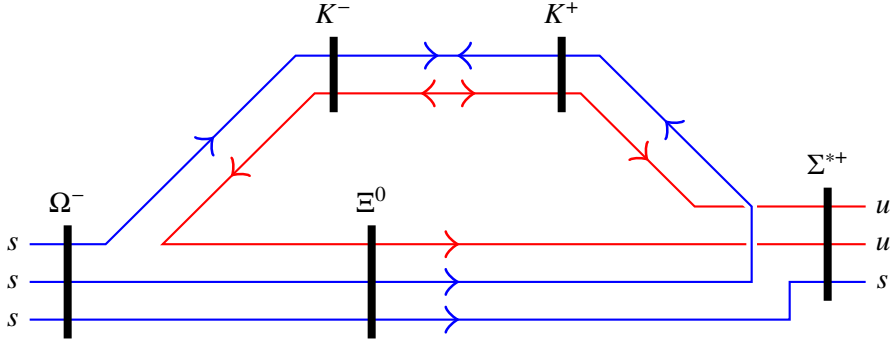
\begin{figure}[th!]
    \centering
\begin{tikzpicture}[scale=0.5]
\draw[red,thick] (22,1) coordinate (H) -- ++(-4.5,0) coordinate (I) -- ++(-3,3) coordinate (L) -- ++(-7,0) coordinate (M) -- ++(-4,-4) coordinate (N) -- (22,0) coordinate (O);

\draw[preaction={draw,line width=4,white},blue,thick] (0,0) coordinate (A) -- ++(2,0) coordinate (B) -- ++(5,5) coordinate (C) -- ++(8,0) coordinate (D) -- ++(4,-4) coordinate (E) -- ++(0,-2) coordinate (F) -- (0,-1) coordinate (G);

\draw[blue,thick] (0,-2) coordinate (P) -- ++(20,0) coordinate (Q) -- ++(0,1) coordinate (R) -- (22,-1) coordinate (S);

\draw[decorate,decoration={markings,mark=at position .57 with {\arrow[blue,scale=3]{>}}}] (B) -- (C);
\draw[decorate,decoration={markings,mark=at position .47 with {\arrow[blue,scale=3]{>}},mark=at position .58 with {\arrow[blue,scale=3]{<}}}] (C) -- (D);
\draw[decorate,decoration={markings,mark=at position .5 with {\arrow[blue,scale=3]{<}}}] (D) -- (E);

\draw[decorate,decoration={markings,mark=at position .5 with {\arrow[red,scale=3]{<}}}] (I) -- (L);
\draw[decorate,decoration={markings,mark=at position .45 with {\arrow[red,scale=3]{<}},mark=at position .6 with {\arrow[red,scale=3]{>}}}] (L) -- (M);
\draw[decorate,decoration={markings,mark=at position .55 with {\arrow[red,scale=3]{>}}}] (M) -- (N);

\draw[decorate,decoration={markings,mark=at position .65 with {\arrow[red,scale=3]{>}}}] (10,0) -- ++(2,0);
\draw[decorate,decoration={markings,mark=at position .65 with {\arrow[blue,scale=3]{>}}}] (10,-1) -- ++(2,0);
\draw[decorate,decoration={markings,mark=at position .65 with {\arrow[blue,scale=3]{>}}}] (10,-2) -- ++(2,0);

\draw[line width=3] (1,-2.5) -- (1,.5) node[above] {$\Omega^-$};
\draw[line width=3] (9,-2.5) -- (9,.5) node[above] {$\Xi^0$};
\draw[line width=3] (8,3.5) -- (8,5.5) node[above] {$K^-$};
\draw[line width=3] (14,3.5) -- (14,5.5) node[above] {$K^+$};
\draw[line width=3] (21,-1.5) -- (21,1.5) node[above] {$\Sigma^{*+}$};

\path (A) node[left] {$s$};
\path (G) node[left] {$s$};
\path (P) node[left] {$s$};

\path (H) node[right] {$u$};
\path (O) node[right] {$u$};
\path (S) node[right] {$s$};
\end{tikzpicture}
    \caption{Example of flavour mixing for $\Omega^{-}$ baryon}
    \label{fig:flavour_mixing}
\end{figure}

A caveat must be made on the ability to determine the spectra of flavoured particles. \Cstar boundary conditions allow some flavour violation when the particles travel around the torus.
Colourless particles only (given a large enough box) may violate flavour under the conditions: $\Delta Q = 0 \text{ mod } 2$; $\Delta B = 0 \text{ mod } 2$; $\Delta F = 0 \text{ mod } 6$ where $Q$ is the electric charge in units of $e$, $B$ is the baryon number and $F = \sum_f F_f$ is the total sum of flavour numbers.
While the flavour mixing of pseudoscalar mesons is harmless as they will not mix with lighter states, and nucleons cannot mix with $B=0$ and are the lightest $B=1$ states, the $\Omega^{-}$ baryons can mix with lighter states. A possible example of this mixing is given in Fig.~\ref{fig:flavour_mixing}. However, it is expected on the basis of a detailed theoretical analysis that any flavour violation is strongly exponentially suppressed with volume~\cite{tantalo_2016} and here we are working under this assumption, postponing as future work a detailed numerical investigation of this issue.

This baryon mass calculation forms part of a larger effort by the RC${}^{*}$ collaboration. These simulations are $1+2+1$ simulations of $O(a)$-improved Wilson fermions with three \Cstar dimensions and periodic boundaries in time. Details of and justification for the trajectories of renormalisation, along which the physical point will be reached, are presented in the companion proceedings~\cite{bushnaq2021update} and will not be discussed here. A Lüscher-Weisz $SU(3)$ gauge action with 
$\beta = 3.24$ is used, along with the SW improvement coefficients depending on the fermion electric charge $c^{q=2/3}_{SW, SU(3)} = c^{q=-1/3}_{SW, SU(3)} = 2.18859$ and $c^{q=2/3}_{SW, U(1)} = c^{q=-1/3}_{SW, U(1)} = 1.0$.
The ensemble analysed in these proceedings is labelled Q*D-$32$-$1$ in the companion proceedings, in which full details of the ensemble may be found. This ensemble, of $64$ time-points and spatial dimensions $L =1.682(5)\text{fm} = 32a$, is relatively far from the physical point, with $\alpha_{\text{em}} = 0.04077(6) \sim 6\alpha_{\text{phys}}$. 

\section{Method}
\subsection{$U(1)$--gauge--invariant charged correlators}
Following the treatment prescribed in Ref.~\cite{tantalo_2016}, we can create through the use of a dressing factor an electrically--charged fermion operator $\Psi$ that is invariant under $U(1)$ local-gauge transformations. We use the `string' dressing factor given in Equation~(3.9) of Ref.~\cite{tantalo_2016}. In the rest of these proceedings, we use capital letters to denote $U(1)$--gauge--invariant quark operators.

\subsection{Baryon interpolating operators} \label{section: bar_interp}

For the proton, which is a spin-$\frac{1}{2}$ baryon, we use the interpolating operator
\begin{equation}
\mathcal{O}^{\pm}_{ a}(x) = P^{\pm}_{ab} \epsilon_{ABC} \left[U^{A}_c(x) \left(\mathcal{C} \gamma_5\right)_{cd} D^B_d (x)\right]  U^C_{b} (x),
\end{equation}
where $A,B,C$ are colour and $a,b,c,d$ are Dirac indices, $\epsilon$ is the anti-symmetric tensor, $U$ and $D$ are the $U(1)$--gauge--invariant up and down quark operators and $\mathcal{C}$ is the charge--conjugation matrix. The neutron operator is simply obtained by $U\leftrightarrow D$. These interpolating operators are known to have a good projection on the ground state in QCD~\cite{zanotti2003}.  The state is projected to a positive or negative parity state using the projector $P^{\pm} = \frac{1}{2} (1 \pm \gamma_0)$. 

The $\Omega^-$ baryon belongs to a vertex of the spin-$\frac{3}{2}$ decuplet and is calculated here using the interpolating operators
\begin{equation}
\mathcal{O}^{i}_{a}(x) =  \epsilon_{ABC} \left[S^{A}_b (x) \left(\mathcal{C} \gamma^i\right)_{bc} S^B_c (x)\right]  S^C_{a} (x),
\end{equation}
where $i$ is a spatial Lorentz index and $S$ is the $U(1)$--gauge--invariant strange quark operator.
The correlator $C^{ij}_{ab}(t) = T\langle 0| \mathcal{O}^i_a(t)\bar{\mathcal{O}}^{j}_b(0)|0\rangle $ contains contributions from spin-$\frac{1}{2}$ and spin-$\frac{3}{2}$ states. Following the treatment in Ref.~\cite{zanotti2003} we use the projection
\begin{equation}
C^{\frac{3}{2}}_{ab} = \sum^3_{i,j=1} \left(C^{ij} \mathcal{P}^{ji}\right)_{ab}; \qquad
\mathcal{P}^{ij} = \delta^{ij} - \gamma^i \gamma^j,
\end{equation}
where $i,j$ are spatial Lorentz indices.
This correlator is then projected to a definite parity state by taking the trace over Dirac indices with the same parity projector $P^{\pm}$ as above:
\begin{equation}
    C^{\frac{3}{2},\pm} = \text{Tr}[C^{\frac{3}{2}} P^{\pm}].
\end{equation}

Both the octet and decuplet correlators are finally folded according to $C(t) = C^+(t)-C^-(T-t)$ to reduce the statistical fluctuations on a given time-slice.

As anticipated in the Introduction, the results presented below correspond to the fermion--connected part of the baryon correlators. Indeed, by relying on the theoretical analysis of Ref.~\cite{tantalo_2016}, we postpone a detailed investigation of the contributions corresponding to the contractions $\langle\Psi(x) \Psi^T(0)\rangle$, that are peculiar of $C^*$--boundary conditions and that induce the spurius flavour mixings discussed above, to future work on the subject.

\subsection{Smearing}
In order to optimise the isolation of the ground state, we use a combination of gradient--flow gauge smearing and Gaussian fermion smearing. We smear the $SU(3)$ gauge fields $V(x,\mu)$ using the gradient--flow specified for periodic spatial boundary conditions~\cite{luscher_2010} by 
\begin{align}
&\dot{V}_t(x, k) = -g_0^2 \{\delta_{x,k}S_w^{\mathrm{spatial}}(V)\}V_t(x,k)\;,  
\qquad  V_0(x,k)=V(x,k)\;,
\qquad  k=1,2,3\;, 
\nonumber \\
&\delta_{x,k}f(V) = T^a \delta^a_{x,k}f(V)\;,  
\qquad\qquad\ \qquad \delta^a_{x,k}f(V) =
 \frac{d}{ds}f(e^{sX}V)\large|_{s=0}\;,
\nonumber \\
&X(y,i) = \begin{dcases} T^a &\mbox{if } (y, i) = (x, k) 
\\ 
0 & \mbox{otherwise}
 \end{dcases} \;,
\end{align}
 to produce smeared gauge links $V_t(x,k)$. Here $T^a$ are the generators of the $SU(3)$ Lie algebra and $S_w^{\mathrm{spatial}}(V)$ is the spatial part of the $SU(3)$--Wilson action (the sum over spatial plaquettes without any prefactor).
 It is important to note here that the smearing is applied on the spatial dimensions only, i.e.~$\dot{V}_t(x,0) = 0$,
 and only on the gauge links that enter into the fermion smearing operators. We have applied to our correlators one level of gauge smearing with evolution time $t = 180 \varepsilon$ with a resolution of $\varepsilon=0.02$.
 We have also checked that this procedure is roughly equivalent to using the more conventional APE smearing when the plaquette, computed in terms of smeared links, is matched. A technical advantage of using the gradient---flow is that unitarity of the links is exactly preserved at any stage.
 
The smeared gauge links are used in the Gaussian smearing of the gauge--invariant fermion operator $\Psi$ to give the smeared operator $\Psi_{\text{smeared}}$:
\begin{align}
&\Psi_{\text{smeared}} = (1+\kappa_{g}H)^{N} \Psi; \\
&H_t(x, y) = \sum^{3}_{j=1}\left\{ V_t(x,j)\delta (x+\hat{j},y)+ V_t(x-\hat{j},j)^{\dagger} \delta (x-\hat{j},y) \right\}\;. 
\end{align}
We have applied Gaussian smearing on both the source and the sink, with three levels of smearing each $N$ = ($0$, $200$, $400$), whilst keeping $\kappa_{g}$ = $0.5$ fixed.

\subsection{Generalised Eigenvalue Problem}
The Generalised Eigenvalue Problem (GEVP)~\cite{gevp} is a standard method of spectral decomposition used to optimise the ground state overlap and to explore excited states. We build a basis of interpolating operators using all possible combinations of three levels of Gaussian smearing on both the source and the sink. All chosen interpolating operators have the same amount of gradient--flow gauge smearing. The correlators with different levels of fermion smearing can then be expressed as a $3$-by-$3$ correlator matrix $C_{nm}$ with $n$ and $m$ indexing the smearing levels on the source and sink respectively. This correlator matrix is then fed into the GEVP. The normalisation time-point for the GEVP was chosen to be $x_0 = 1$. 

\begin{figure}
    \centering
    \begin{subfigure}[t]{0.5\textwidth}
    \centering
    \includegraphics[width=\textwidth]{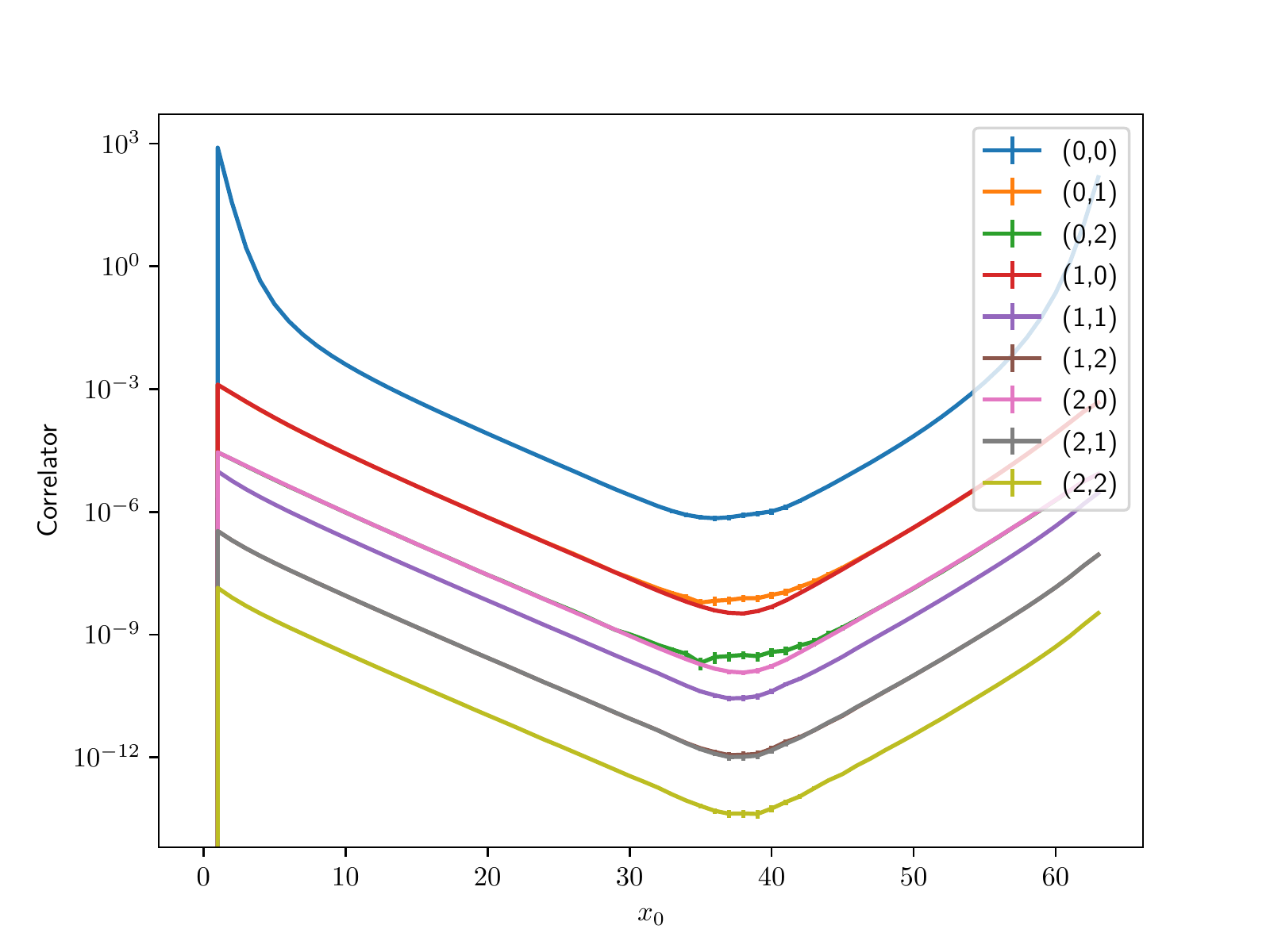}
    \caption{Correlator for different levels of Gaussian smearing} \label{fig: n_corr}
    \end{subfigure}\hfill
    \begin{subfigure}[t]{0.5\textwidth}
    \centering
    \includegraphics[width=\textwidth]{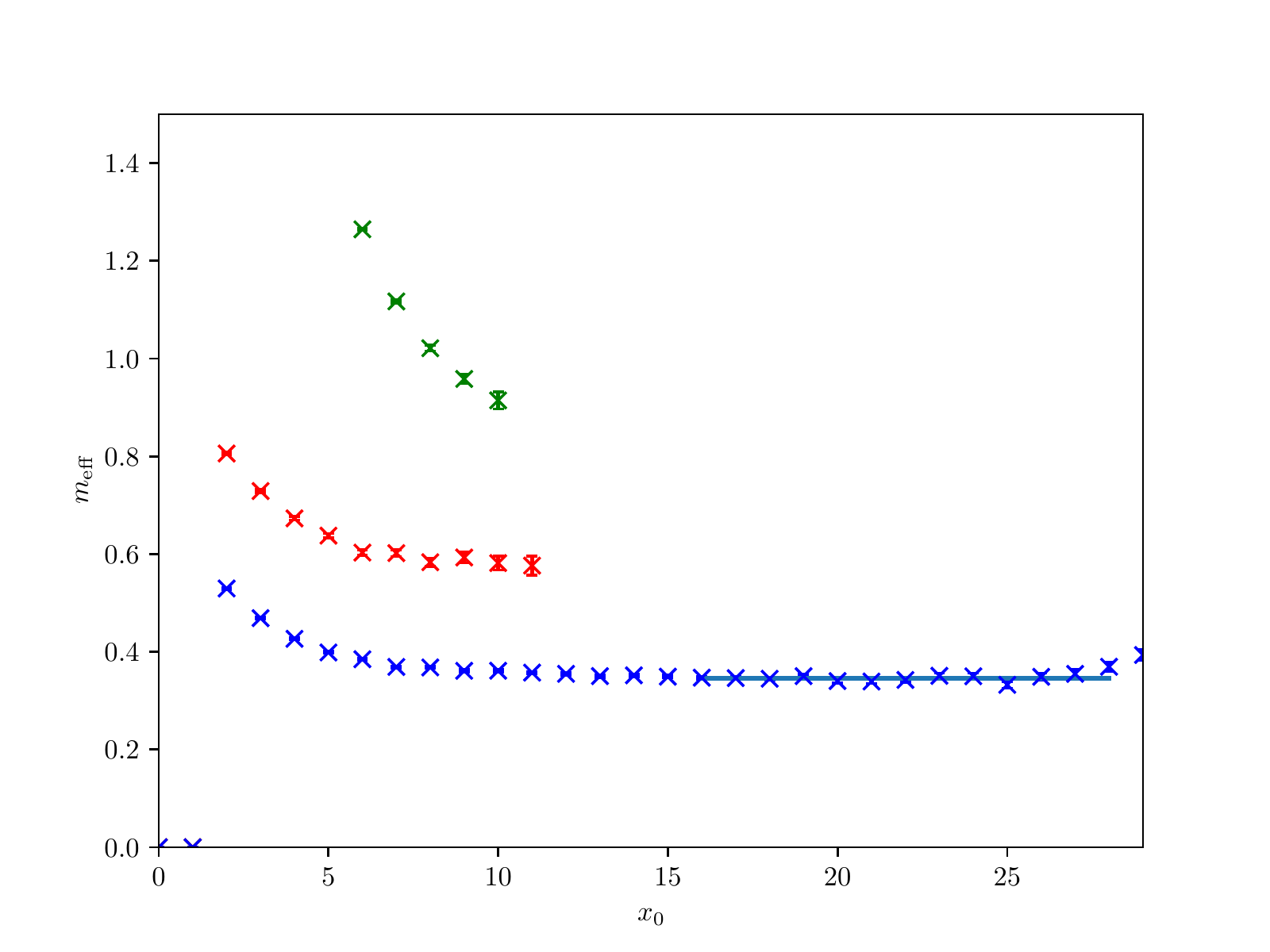}
    \caption{Spectrum from GEVP} \label{fig: n_gevp}
    \end{subfigure}
    \caption{\textbf{Neutron analysis.} Fig.~\ref{fig: n_corr} shows the neutron correlator at three different levels of fermion smearing at both the source and the sink; the correlators are labelled in the legend as $(n_{\text{source}},n_{\text{sink}})$, indicating $n_{\text{source}}$ smearing levels on the source and $n_{\text{sink}}$ levels on the sink.  We see that increased smearing reduces the curvature of the correlator at small times. These correlators with different smearing levels when analysed using the GEVP give three distinct energy levels, as shown in Fig.~\ref{fig: n_gevp}, with the ground state in blue, for which we see a long plateau, and excited states in red and green. The plateau without error is shown in lighter blue, with the range of the line showing the points used in the fit.}
    \label{fig:neutron}
\end{figure}

Fig.~\ref{fig: n_corr} shows the neutron correlator for different levels of Gaussian smearing, folded as described in Section~\ref{section: bar_interp}, and then solved with the GEVP to obtain the spectrum, see Fig.~\ref{fig: n_gevp}. 

\section{Results}

\begin{figure}[ht!]
    \centering
    \begin{subfigure}[t]{0.5\textwidth}
    \centering
    \includegraphics[width=\textwidth]{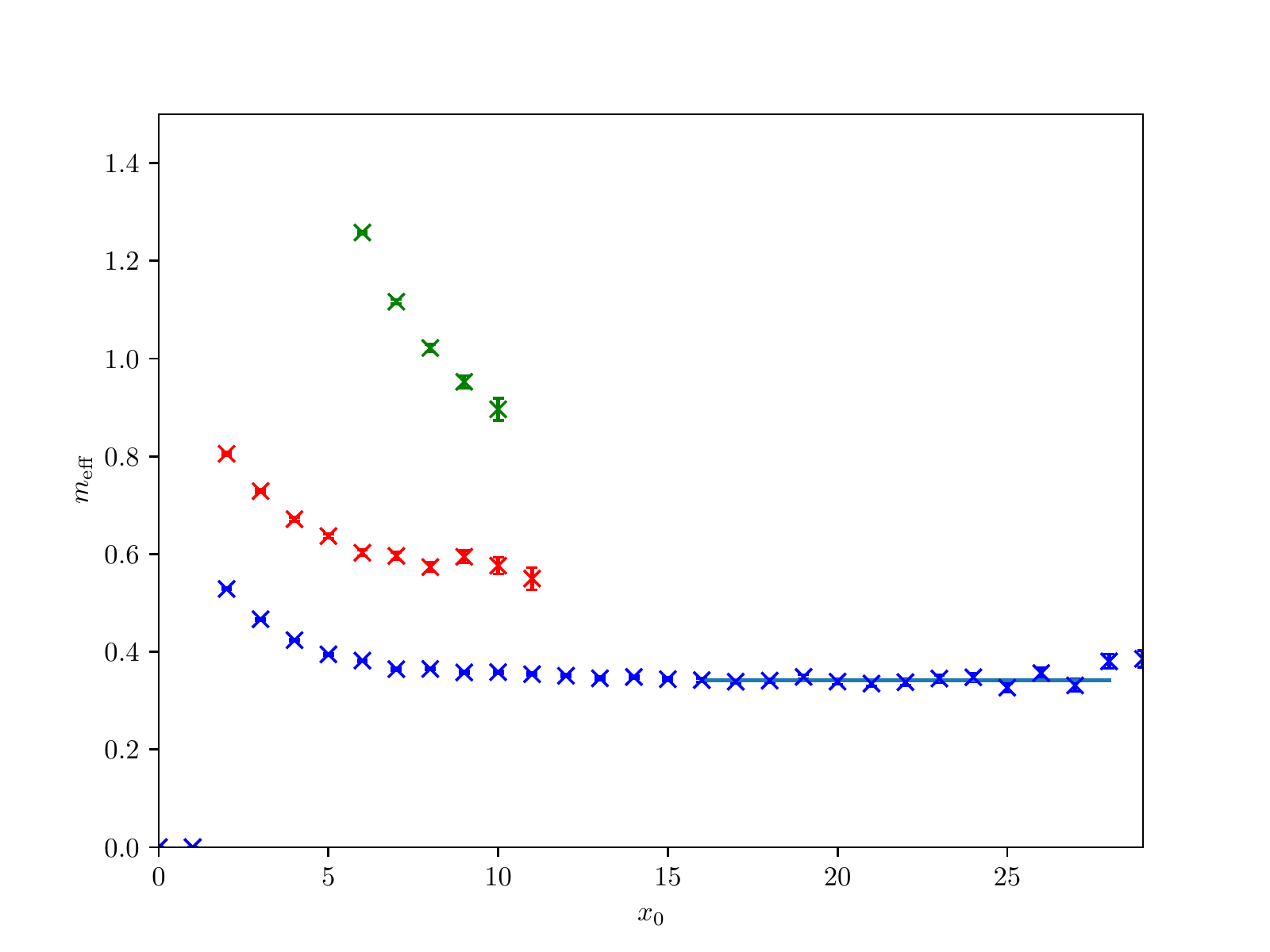}
    \caption{Proton spectrum}\label{fig:proton_spectrum}
    \end{subfigure}\hfill
    \begin{subfigure}[t]{0.5\textwidth}
    \centering
    \includegraphics[width=\textwidth]{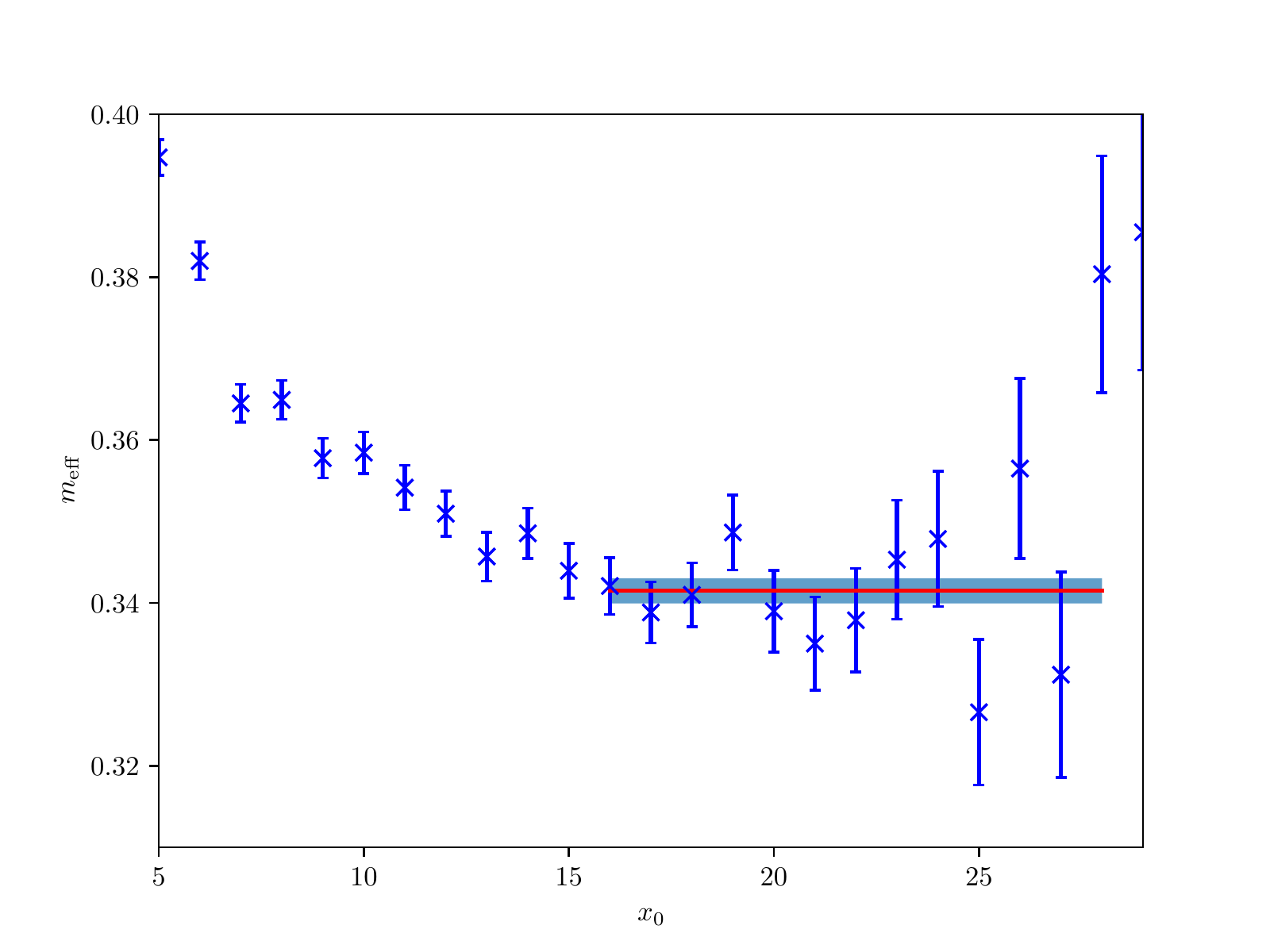}
    \caption{Proton ground state plateau}\label{fig:proton_plateau}
    \end{subfigure}
    \caption{\textbf{Proton analysis.} Fig.~\ref{fig:proton_spectrum} shows the proton spectrum from the GEVP. The ground state effective mass is given in blue, while excited states are shown in red and green. The ground state is shown in more detail in Fig.~\ref{fig:proton_plateau}, with the plateau in red and its error in lighter blue. We obtain a mass of $m_p = 1282(8)$~MeV.}
    \label{fig:proton}
\end{figure}

\begin{figure}[ht!]
    \centering
    \begin{subfigure}{0.5\textwidth}
    \centering
    \includegraphics[width=\textwidth]{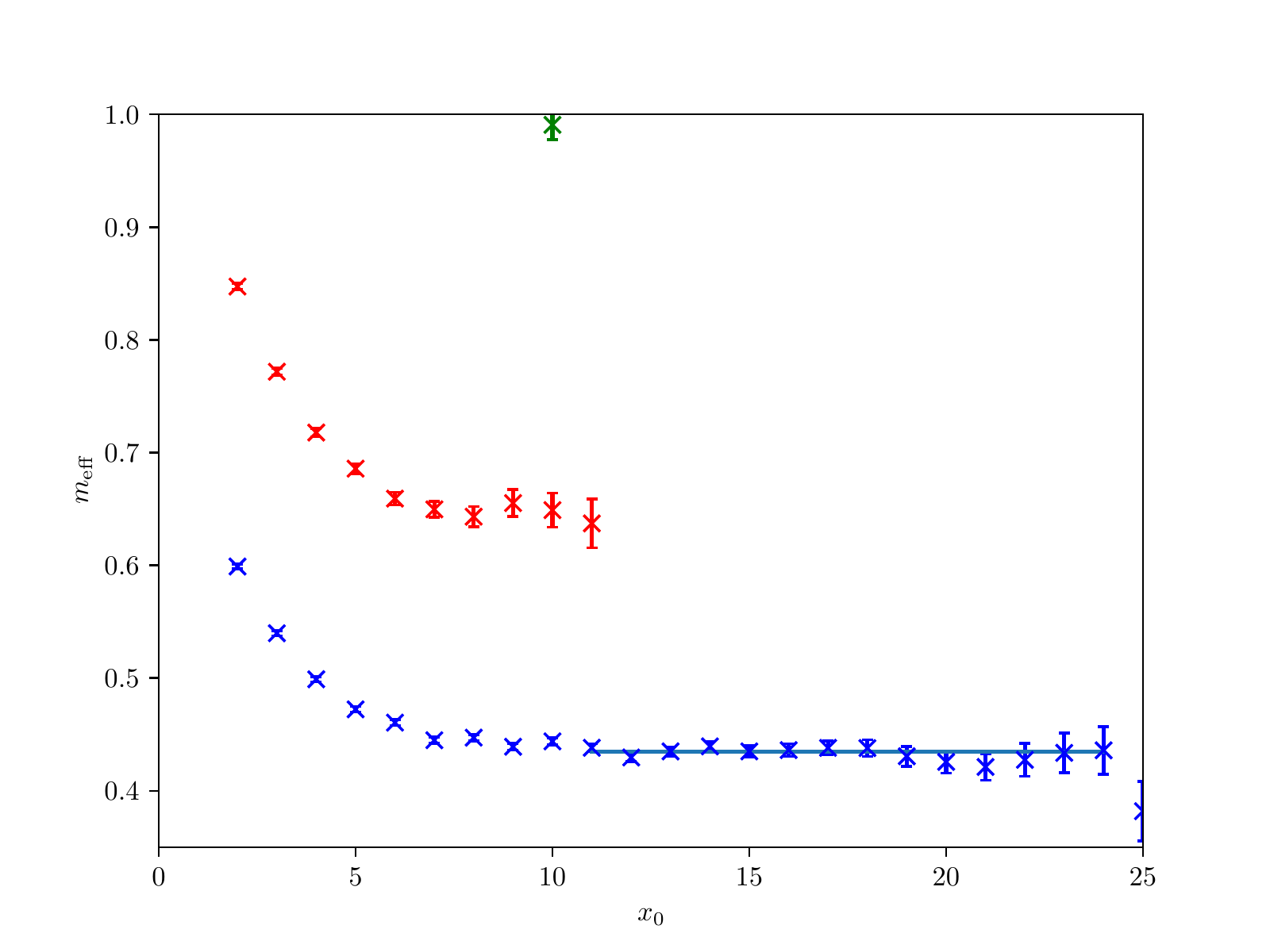}
    \caption{$\Omega^{-}$ spectrum}\label{fig:omega_spectrum}
    \end{subfigure}\hfill
    \begin{subfigure}{0.5\textwidth}
    \centering
    \includegraphics[width=\textwidth]{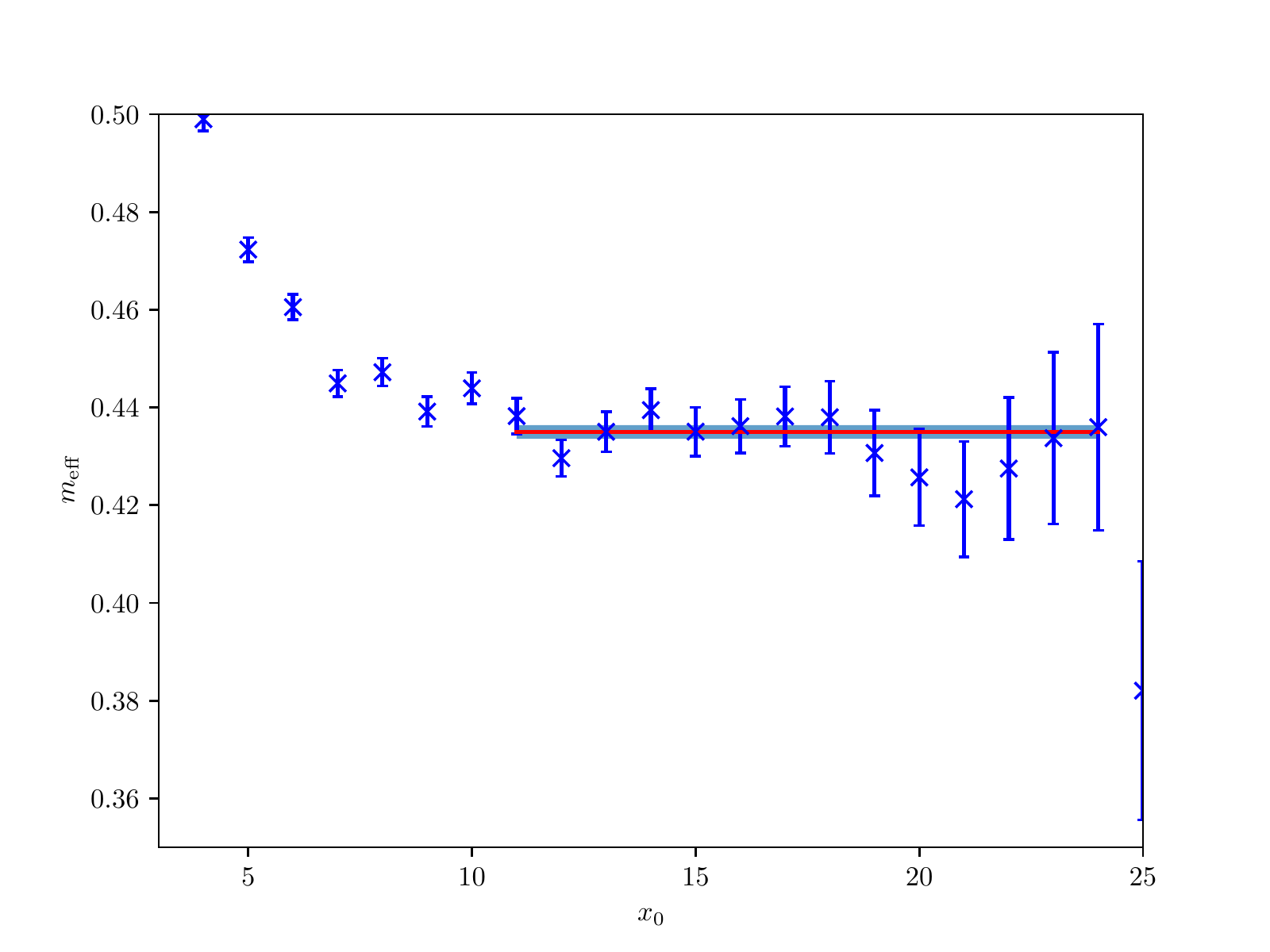}
    \caption{$\Omega^{-}$ ground state plateau}\label{fig:omega_plateau}
    \end{subfigure}
    \caption{\textbf{$\Omega^{-}$ baryon analysis.} Fig.~\ref{fig:omega_spectrum} shows the GEVP output for the $\Omega^{-}$ baryon. The blue points show the ground state effective mass, whereas the red and green points show the excited states. Fig.~\ref{fig:omega_plateau} shows the ground state in more detail. The plateau is shown in red and its error in lighter blue, and we see that the plateau is rather long for these statistics. We find a mass of $m_{\Omega^{-}} = 1633(8)$~MeV.}
    \label{fig:omega}
\end{figure}

\begin{figure}[ht!]
    \centering
    \includegraphics[width=0.5\textwidth]{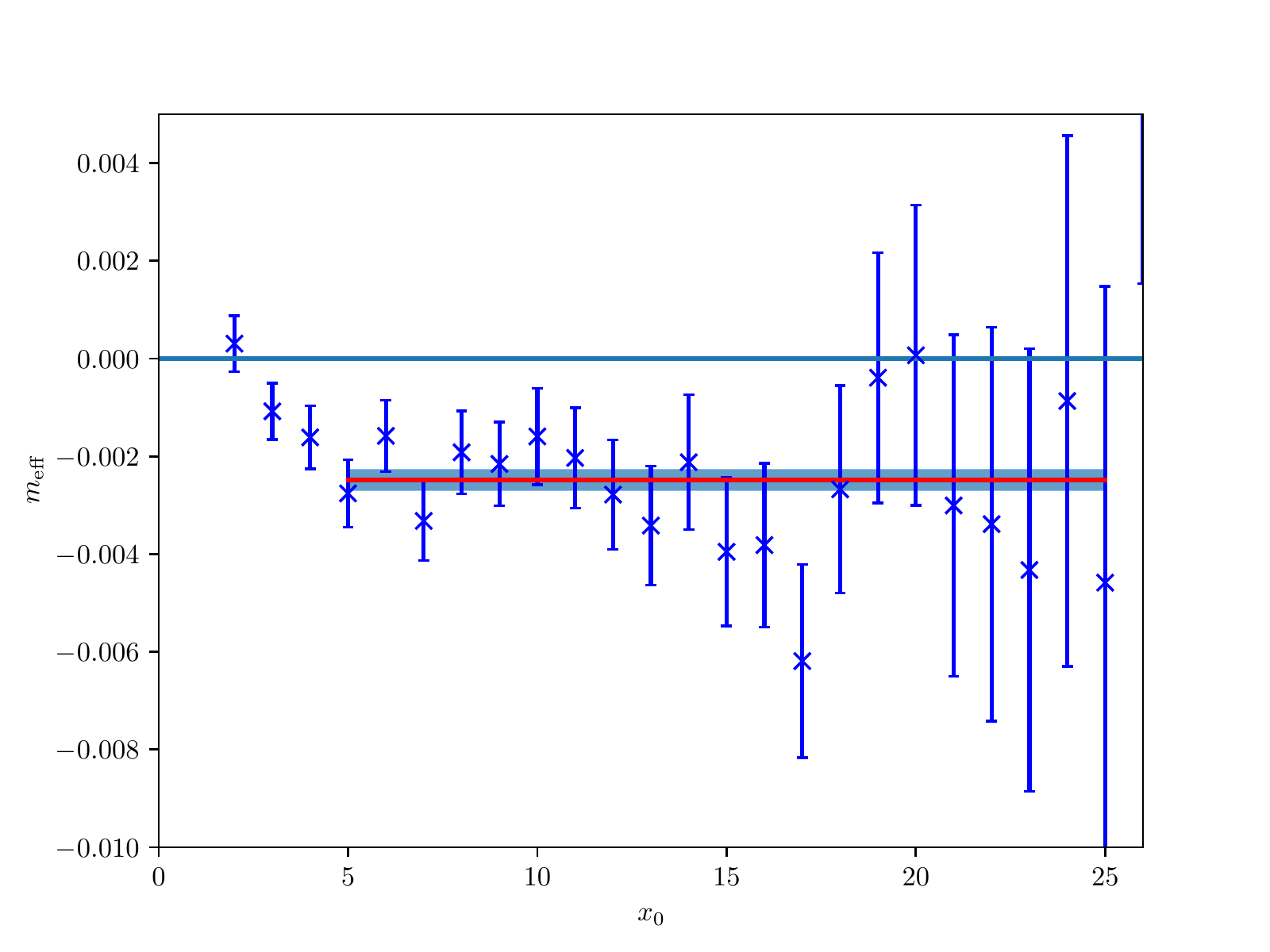}
    \caption{\textbf{Proton-neutron mass difference analysis.} The difference between the proton and neutron effective mass curves is shown here. The plateau is shown in red and its error is shown in lighter blue, with the zero-intercept also shown in blue. We see a plateau that starts at a small $x_0$-value and that the signal lasts for a reasonably long $x_0$-length, giving the mass difference $m_n - m_p = 9(1)$~MeV.}
    \label{fig: pn_md}
\end{figure}

The results presented here were obtained with $1993$ gauge configurations by performing four point--like propagator inversions starting from random points. We stress once again that this ensemble is rather far from the physical point. In order to help the interpretation of the baryon mass results given below we quote for reference the value of the charged pion (and kaon) mass calculated for this ensemble, $m_{\pi^+}=m_{K^+}= 496(2)$~MeV (see Ref.~\cite{bushnaq2021update} for more details). 
This result, as well as all the following ones, take into account the \textit{universal} finite volume corrections on charged hadron masses given by the first two terms of Eq.~($5.1$) in Ref.~\cite{tantalo_2016}.

The proton spectrum is shown in Fig.~\ref{fig:proton_spectrum}. We see a reasonably long ground state plateau for the statistics used, with a clear first excited state. The ground state plateau and its error are shown more clearly in Fig.~\ref{fig:proton_plateau}. The mass of the ground state is found to be $1282(8)$~MeV. An estimate of the lowest energy gap was found to be roughly consistent with a proton + photon state.

Next, we present the $\Omega^-$ baryon in Fig.~\ref{fig:omega}. We find a plateau that starts early and persists until around $x_0=24$. The mass result we obtain is $1633(8)$~MeV. The energy gap was found to be roughly equivalent to an $\Omega^-$ + photon state.

Fig.~\ref{fig: pn_md} shows the mass difference between the proton ground state and the neutron ground state, i.e.~the proton-neutron mass difference. We see here a plateau that starts early but the signal-to-noise ratio becomes poor around $x_0=20$. We find the mass difference to be $m_n - m_p = 9(1)$~MeV. It is worth noting here once again the un-physicality of the ensemble if one is tempted to compare it to the physical value of roughly $1$~MeV.

\section{Conclusion}
The results on the baryon mass spectrum presented in these proceedings form part of a larger effort by the RC$^{*}$ collaboration. The ultimate goal of this effort is to obtain physical results for the hadron spectrum by performing first--principles lattice simulations of QCD$+$QED without relying at any stage on gauge--fixing or perturbation theory. Alongside the companion proceedings~\cite{bushnaq2021update}, a first step towards this goal has been made here with results obtained on a single ensemble of gauge configurations corresponding to an unphysical setup in which $\alpha_{\text{em}}\simeq 6 \alpha_{\text{phys}}$, the lattice volume is $L\simeq 1.7$~fm with a lattice spacing $a\simeq 0.05$~fm and the four dynamical quark masses have been tuned at the $U$--spin symmetric point $m_d=m_s$.

Our results demonstrate that, in addition to charged meson masses, baryon masses can be calculated  with satisfactory precision in QCD$+$QED$_C$ in a fully local and gauge--invariant setup. This makes us reasonably confident in the possibility of reaching our goal and providing phenomenologically relevant results on the full hadron spectrum in the near future.

\acknowledgments
This project has received funding from the European Union's Horizon 2020 research and innovation programme under grant agreement No 765048.
 The research of AC, JL and AP is funded
by the Deutsche Forschungsgemeinschaft (DFG, German Research Foundation) - Projektnummer
417533893/GRK2575 “Rethinking Quantum Field Theory”. 
The authors acknowledge access to the Eagle HPC cluster at PSNC (Poland).
The work was supported by the Poznan Supercomputing and Networking Center (PSNC) through grants 450 and 466.
The work was supported by CINECA that granted computing resources on the Marconi supercomputer to the LQCD123 INFN theoretical initiative under the CINECA-INFN agreement.
We acknowledge access to Piz Daint at the Swiss National Supercomputing Centre, Switzerland under the ETHZ's share with the project IDs go22 and go24.
The work was supported by the North-German Supercomputing Alliance (HLRN) with the project bep00085.

\bibliographystyle{JHEP}
\bibliography{sample.bib}

\providecommand{\href}[2]{#2}\begingroup\raggedright\begin{thebibliography}{1}

\bibitem{bushnaq2021update}
L.~Bushnaq, I.~Campos, M.~Catillo, A.~Cotellucci, M.~Dale, P.~Fritzsch et~al.,
  \emph{An update on {QCD+QED} simulations with {$C^{*}$} boundary conditions},
   \href{https://arxiv.org/abs/2108.11989}{{\ttfamily 2108.11989}}.

\bibitem{bmw_2015}
S.~Borsanyi, S.~Durr, Z.~Fodor, C.~Hoelbling, S.D.~Katz, S.~Krieg et~al.,
  \emph{Ab-initio calculation of the neutron-proton mass difference},
  \href{https://doi.org/10.1126/science.1257050}{\emph{Science} {\bfseries 347}
  (2015) 1452–1455}.

\bibitem{kronfeld1991}
A.~Kronfeld and U.~Wiese, \emph{{SU(N)} gauge theories with {C}-periodic
  boundary conditions ({I}). {T}opological structure},
  \href{https://doi.org/https://doi.org/10.1016/0550-3213(91)90479-H}{\emph{Nuclear
  Physics B} {\bfseries 357} (1991) 521}.

\bibitem{tantalo_2016}
B.~Lucini, A.~Patella, A.~Ramos and N.~Tantalo, \emph{Charged hadrons in local
  finite-volume {QED+QCD} with {$C^{*}$} boundary conditions},
  \href{https://doi.org/10.1007/jhep02(2016)076}{\emph{Journal of High Energy
  Physics} {\bfseries 2016} (2016) }.

\bibitem{zanotti2003}
J.M.~Zanotti, D.B.~Leinweber, A.G.~Williams, J.B.~Zhang, W.~Melnitchouk and
  S.~Choe, \emph{Spin-3/2 nucleon and {$\Delta$} baryons in lattice {QCD}},
  \href{https://doi.org/10.1103/physrevd.68.054506}{\emph{Physical Review D}
  {\bfseries 68} (2003) }.

\bibitem{luscher_2010}
M.~Lüscher, \emph{Properties and uses of the {W}ilson flow in lattice {QCD}},
  \href{https://doi.org/10.1007/jhep08(2010)071}{\emph{Journal of High Energy
  Physics} {\bfseries 2010} (2010) }.

\bibitem{gevp}
B.~Blossier, M.D.~Morte, G.~von Hippel, T.~Mendes and R.~Sommer, \emph{On the
  generalized eigenvalue method for energies and matrix elements in lattice
  field theory},
  \href{https://doi.org/10.1088/1126-6708/2009/04/094}{\emph{Journal of High
  Energy Physics} {\bfseries 2009} (2009) 094}.

\end{thebibliography}\endgroup

\end{document}